\documentclass[aps,prd,twocolumn,showkeys,superscriptaddress,singlecolumn,nofootinbib,floatfix]{revtex4-2}
\usepackage{amssymb,amsmath}
\DeclareMathOperator\arctanh{arctanh}
\usepackage{datetime}
\usepackage{color}
\usepackage{graphicx}
\usepackage{hyperref}
\usepackage{algpseudocode}
\usepackage[]{mdframed}
\usepackage{enumitem}
\usepackage{pifont}
\usepackage{cancel}
\usepackage{ulem}
\begin{document}
\author{Gustavo Godinho}
\email{\textcolor{black}{gustavo.breno@aluno.ufabc.edu.br}}
\affiliation{Centro de Ciências Naturais e Humanas, Universidade Federal do ABC,
Av. dos Estados 5001, 09210-580 Santo André, São Paulo, Brazil}
\author{Willians Barreto}
\email{\textcolor{black}{willians.barreto@ufabc.edu.br}}
\affiliation{Centro de Ciências Naturais e Humanas, Universidade Federal do ABC,
Av. dos Estados 5001, 09210-580 Santo André, São Paulo, Brazil}
\affiliation{Centro de Física Fundamental, Universidad de Los Andes, Mérida 5101, Venezuela}
\title{\textcolor{black}{Are these quasi-normal modes?\\{\it \textcolor{black}{Esses são modos quase-normais?}}}}

\begin{abstract}  
\textcolor{black}{We discuss how to extract numerically the expected lowest quasi-normal mode (QNM) associated with the pressure anisotropy for a Bjorken flux evolution to equilibrium. This QNM was easily decoded subtracting the hydrodynamical attractors and compared with other authors calculations. After evolutions with transients close to the expected lowest QNM the system goes to a tail (pure imaginary frequency) for late times. We analyze the relevance of Navier-Stokes, second order and Borel attractors at each stage of the evolution, which begins far-from-equilibrium and ends close to equilibrium.\\
{\it \textcolor{black}{Discutimos como extrair numericamente o modo quase-normal (MQN) esperado mais baixo associado à pressão anisotrópica para um fluxo de Bjorken evoluindo para o equilíbrio. Este MQN foi facilmente decodificado subtraindo os atratores hidrodinâmicos e comparado com os calculados por outros autores.   Após evoluções com transientes próximos ao MQN mais baixo esperado, o sistema vai para uma cauda (frequência imaginária pura) para tempos tardios. Analisamos a relevância dos atratores de Navier-Stokes, de segunda ordem e de Borel em cada estágio da evolução, que começa longe do equilíbrio e termina perto do equilíbrio.}}}  
\end{abstract}
\date{\today, [\currenttime]}
\maketitle
\textcolor{black}{\section{Introduction}}
\textcolor{black}{The main purpose of this work is to show how we can extract extra physical information from an available and yet squeezed data. What a naked eye does not see, and from a naïve (fresh) point of view. In current times a huge quantity of certified data is in the cloud which can be analyzed using Artificial Intelligence (AI) task-force algorithms, for example. That is not the case here, but we will keep it in mind for future work based on the our results. Instead we will implement a very simple idea to go deep into the data and learn.}

\textcolor{black}{The context is the quark-gluon plasma (QGP) formed in ultrarelativistic heavy-ion collisions \cite{hs13}, to study out of equilibrium properties \cite{bhmv21}.  \textcolor{black}{How does hydrodynamics emerge from the non-equilibrium regime?} The current view is that hydrodynamics is a universal attractor \cite{hs15}, with dissipative contributions even when local gradients are large \cite{fhs18}.}  
\textcolor{black}{In the simulation of a QGP the hydrodynamics can be modeled using kinetic theory \cite{kl14}, \cite{kmpst19}, \cite{aks20} and holographic duality \cite{m98}, \cite{gkp98}, \cite{w98a}, \cite{w98b}. Kinetic descriptions show purely exponential decay of nonhydrodynamic modes \cite{dnnr11}, while in holography these modes also feature an oscillatory behavior \cite{fhs18}, \cite{ks05}, \cite{hjsw14}.}

\textcolor{black}{From a practical point of view, our input data is the output of a holographic numerical model studied recently in \cite{rnbdd21}, \cite{rbn22}, \cite{rb22}, for the simplest model. That is the case of a Supersymmetric Yang-Mills (SYM) \textcolor{black}{plasma} undergoing a Bjorken flow \cite{b83}. Our motivation was a recent work on homogeneous isotropization \cite{rb24}, with non-hydrodynamical homogeneous {QNMs}\footnote{\textcolor{black}{QNMs are the response of black holes to perturbations in different channels; are damped oscillations characterized by complex eigenfrequencies.}} once they trespass some threshold to thermalize, correlated with a stairway to equilibrium entropy. In a general initial setting out of equilibrium, the Bjorken flux clearly first goes to hydrodynamization and then to thermalization, with a transient violation of the energy conditions in some cases. The oscillatory behavior in the pressure anisotropy is observed particularly strong when the energy conditions are transiently violated to finally go to an apparent exponential decay. Are those oscillatory features related with QNMs? Why the oscillations decay to a tail for all evolved initial data? Why plateaus in entropy are formed for the Bjorken flow? Why the stairway to equilibrium entropy is not observed for the Bjorken flow? Here we try to give some answers or at least report credible evidence, analyzing results produced with the available numerical code of a known precision.} 

\textcolor{black}{We organize this work as follows. For the sake of completeness we present in section II the most salient features of the studied model (including initial conditions), the holographic description of the Bjorken flux, a general notion of hydrodynamic attractors, as for a non expert reader but interested in the field. In section III we describe the numerical code developed as the robot used in this work. We present our numerical results in section IV to finally discuss them and conclude in section V.}

\textcolor{black}{Here we use plus metric signature and natural units $\hbar = c = k_B = 1$.}
\vspace{0.5cm}
\textcolor{black}{\section{Holographic Bjorken flow}\label{II}}
\textcolor{black}{\subsection{Model}\label{II.A}}
\textcolor{black}{A foundational model to study the matter created in relativistic nucleus--nucleus collisions is the Bjorken flow \cite{b83}. In this model the relativistic fluid possesses the following features (symmetries). Taking the collision axis to be the longitudinal $z$-axis, it is assumed boost invariance in that direction where the fluid is inhomogeneous and rapidly expanding at the speed of light. In the transverse $xy$ plane the nuclei is assumed to be homogeneous\footnote{\textcolor{black}{Transverse distances are considered much smaller than the nuclear radii.}} and of infinite extent, removing thus all the dependence on the coordinates $x$ and $y$ \cite{rr19}. 
Bjorken symmetry is more easily handled by changing from Cartesian coordinates $(t, x, y, z)$ to the so-called Milne coordinates $(\tau,x,y,\xi)$, where $\tau=\sqrt{t^2-z^2}$ is the proper time and 
$\xi=\tfrac{1}{2}\ln\{(z+t)/(z-t)\}=\arctanh(z/t)$ is the spacetime rapidity. In these coordinates all the hydrodynamic fields are exclusive functions of the proper time\footnote{\textcolor{black}{The fluid expanding longitudinally in the static Minkowski spacetime is equivalent to at rest fluid in the longitudinally expanding (flat) spacetime. }} $\tau$. 
In terms of the Milne coordinates the metric of the 4D Minkowski spacetime where the fluid is defined reads,
\begin{align}
ds^2_{\textrm{(4D)}} = -d\tau^2+dx^2+dy^2+\tau^2 d\xi^2.
\label{Minkowski}
\end{align}
This setting is a first approximation to the expanding quark-gluon plasma (QGP) formed in high-energy heavy-ion collisions near mid-rapidity, i.e. close to the collision axis (the transverse expansion to the collision axis is completely neglected). This basic model can be extended to consider an anisotropic (viscous) holographic fluid \cite{s18} (see also  \cite{rnbdd21}, \cite{rbn22} and references therein).}

\textcolor{black}{\subsection{Correspondence}}

\textcolor{black}{
The holographic Bjorken flow of a relativistic and strongly coupled quantum fluid can be implemented by considering that the 4D flat spacetime (\ref{Minkowski}), where the fluid lives, is (up to a global conformal factor) the boundary of a 5D curved spacetime asymptotically anti-de Sitter, the bulk. The simplest holographic model is the conformal and strongly coupled SYM plasma. In some limit the gauge/gravity duality states that the SYM in 4D Minkowski spacetime is dual to classical (Einstein) gravity in 5D.}

\textcolor{black}{The Ansatz for the 5D bulk metric can be written using infalling Eddington-Finkelstein (EF) coordinates as follows \cite{cy10}, \cite{cy13}
\begin{align}
ds^2 =& \,2d\tau\left[dr-A(\tau,r) d\tau \right]+\Sigma(\tau,r)^2\left[e^{-2B(\tau,r)}d\xi^2 \right.\nonumber\\ 
      & \left. +\, e^{B(\tau,r)}(dx^2+dy^2)\right],
\label{lineElement}
\end{align}
where $r$ is the radial holographic direction, $\tau$ is the EF time which reduces to the proper time in (\ref{Minkowski}) at the boundary ($r\rightarrow\infty$), $x$ and $y$ are the coordinates in the plane transverse to the beamline and $\xi$ is the rapidity in the longitudinal direction, as described in section \ref{II.A}. In such a way
\begin{equation}
\lim_{r\rightarrow\infty}ds^2=ds^2_{(4D),}
\end{equation}
leads to the following boundary conditions
\begin{subequations}
\begin{align}
\lim_{r\rightarrow\infty} A &\rightarrow \tfrac{1}{2}r^2,\\
\lim_{r\rightarrow\infty} B &\rightarrow -\tfrac{2}{3}\ln\tau,\\
\lim_{r\rightarrow\infty} \Sigma &\rightarrow r\tau^{1/3}.
\end{align}
\label{bc}
\end{subequations}
In the EF coordinates infalling radial null geodesics satisfy $\tau=$ constant, while outgoing radial null geodesics satisfy $dr=Ad\tau$. By foliating the bulk spacetime in slices of constant $\tau$ (null hypersurfaces)   
one implements a time evolution of the system according to the so-called characteristic formulation of general relativity 
(for a detailed review including numerical issues see \cite{cy13}). For the original characteristic formulation involving 
asymptotically flat spacetimes see \cite{bondi}, \cite{sachs} (and for a practical review see \cite{winicour}).} 

\textcolor{black}{The line element (\ref{lineElement}) still has a residual diffeomorphism invariance under radial shifts, that is, $r \mapsto r + \lambda(\tau)$, with $\lambda(\tau)$ an arbitrary function of time. In order to integrate in the radial direction the Einstein equations, one must consider the entire portion of the bulk geometry causally connected to the boundary. Using the residual diffeomorphism invariance we can deal with black holes, requiring that the radial position of the apparent horizon remains fixed for all the time slices.}

\textcolor{black}{Einstein's equations are obtained --straight forward-- from (\ref{lineElement}) and solved
numerically considering the boundary conditions (\ref{bc}) and for arbitrary initial conditions. This was routinely done in characteristic numerical relativity \cite{winicour}.}

\textcolor{black}{What it is most interesting to consider is the ultraviolet (UV) near boundary expansions of the metric coefficients (reminiscent of the Bondi formalism to deal with gravi\-tational radiation \cite{bondi}), and their relation to the holographically normalized one-point Green's function of the energy momentum tensor of the boundary SYM gauge theory. From this procedure we extract the physical observables of the strongly coupled fluid under consideration. Thus, the minimalist metric fields reads:
\begin{subequations}
\begin{align}
A(\tau,r)&= \tfrac{1}{2}[r+\lambda(\tau)]^2+\dot\lambda(\tau)+\frac{a_2(\tau)}{r^2}+\mathcal{O}(r^{-3}),\\
B(\tau,r) &=-\tfrac{2}{3}\ln\tau + \mathcal{O}(r^{-1}),\\
\Sigma(\tau,r) &=\tau^{1/3}r + \frac{1+3\tau\lambda(\tau)}{3\tau^{2/3}} + \mathcal{O}(r^{-1}),
\end{align}
\label{expans}
\end{subequations}\\
where $\lambda(\tau)$ and $a_2(\tau)$ have to be specified at $\tau=\tau_0$. After the holographic correspondence task, are obtained the following formulas for observables:
\begin{subequations}
\begin{align}
\hat\epsilon(\tau)&= -3a_2(\tau),\\
\hat p_T\tau) &=-3a_2(\tau)-\frac{3}{2}\tau\dot a_2(\tau),\\
\hat p_L(\tau) &= 3a_2(\tau)+3\tau\dot a_2(\tau),
\end{align}
\label{expans}
\end{subequations}\\
where $\hat\epsilon(\tau)$, $p_T(\tau)$, $p_L(\tau)$ are, respectively, the (normalized) energy density, the transverse pressure, and the longitudinal pressure of the SYM plasma. Note that an overdot represents a time derivative with respect to time. Once determined the time evolution of $a_2(\tau)$ UV coefficient we have the dynamical evolution of the physical observables at the boundary\footnote{\textcolor{black}{In some way the holographic correspondence can be considered an algorithm, with the universe as a classical computer doing a simulation of a dual quantum computer; the played role of a black hole is fundamental for thermalization and for the holographic realization itself.}}.} 
\vspace{0.5cm}
\textcolor{black}{\subsection{Attractors}}
\textcolor{black}{A Bjorken flow is an ideal playground to test the current notion of hydrodynamics: attractors. They drive in some stage the behavior for observables such as the pressure anisotropy, which is dictated by the decay of the nonhydrodynamic QNM. As proposed in \cite{hs15} the Borel resummation of the divergent asymptotic gradient series defines a hydrodynamic atrractor, to which far-from-equilibrium solutions would coalesce before converging to the corresponding limits associated with finite order truncations of the hydrodynamic gradient expansion, such as Navier-Stokes theory or the second (or higher) order hydrodynamic  \cite{rbn22}.}

\textcolor{black}{For the pressure anisotropy of the SYM plasma undergoing a Bjorken flow, the corresponding analytical hydrodynamic expressions for the Navier-Stokes (NS) regime, the second-order gradient expansion, and the Borel resummation of the divergent gradient expansion are given by, respectively \cite{s18}, \cite{r18}, \cite{brsss08}, \cite{hjw13}
\begin{subequations}
\begin{align}
\left[\frac{\Delta\hat p}{\hat\epsilon}\right]_{\text{NS}}&=\frac{2}{3\pi\omega_\Lambda},\label{NS}\\
\left[\frac{\Delta\hat p}{\hat\epsilon}\right]_{\text{2nd order}}&=\frac{2}{3\pi\omega_\Lambda} +
\frac{2(1-\ln(2))}{9\pi^2\omega^2_\Lambda} \label{2nd},\\
\left[\frac{\Delta\hat p}{\hat\epsilon}\right]_{\text{Borel}}&=\frac{-276+2530\omega_\Lambda}{3(120-570\omega_{\Lambda}+3975\omega_{\Lambda}^2)} \label{borel}.
\end{align}
\label{attractors}
\end{subequations}\\
with $\Delta\hat p\equiv \hat p_T-\hat p_L$ defined as the pressure anisotropy and $\omega_\Lambda$ the dimensionless time measure, scaled by the energy parameter $\Lambda$ (as fixed in \cite{crn19}) which in turn depends on each initial condition. These expressions are interpreted as hydrodynamic attractors \cite{rbn22}, adapted to our holographic model.}

\textcolor{black}{\subsection{Initial conditions}}
\textcolor{black}{In order to prepare a code to extract the observables we require define the compactified holographic coordinate
\begin{equation}
u\equiv\frac{1}{r},
\end{equation}
in terms of which are defined the subtracted\footnote{\textcolor{black}{This is a common practice in characteristic numerical relativity \cite{winicour}}.} field variables:
\begin{equation}
u^pX_s(\tau,u)\equiv X(\tau,u)-X_{UV}(\tau,u),
\end{equation}
where $X$ denotes any of the metric functions, $p$ is an integer, and $X_{UV}$ is some truncation of $X$.
At this point, and without loss of generality, we set $\lambda(\tau)=0$ for any time and for this work purposes.
Thus we use here the set of initial conditions (IC) for $B_s$ and $a_2$ as reported in \cite{rnbdd21} by Table I (also used in \cite{rbn22} and \cite{kaminski}})

\hspace{.5cm}

\textcolor{black}{\section{Automate codes}}
\textcolor{black}{The numerical code used in \cite{rnbdd21}, \cite{rbn22} was automatized for this work. The original code implemented the pseudo-spectral method to solve the characteristic hypersurface equations, and the Adams-Bathforth method to evolve the initial/boundary conditions. 
The features of the robot written in Python/Fortran are:\\
\begin{mdframed}
\textcolor{black}{
\begin{itemize}
\item {\sc Input}: {\it Read a set of parameters for any initial condition IC};
\item {\sc Input}: {\it Read a set of ICs};
\item {\sc Wrapper: \\{\it For each IC (serial or parallel)}}:
\begin{itemize}[label=\ding{212}]
\item {\sc Driver}: {\it Integrate the Einstein field equations up to some time};
\begin{itemize}[label=$\checkmark$]
\item {\sc Output}: {\it  Write data at each time slice};
\end{itemize}
\item {\sc Post-processing}: {\it Calculate the energy parameter} $\Lambda$;
\begin{itemize}[label=$\checkmark$]
\item {\sc Input}: {\it Read the {\sc output} and adjust observables and time};
\item {\sc Output}: {\it Write the scaled data};
\item {\sc Output}: {\it Generate and save plots};
\end{itemize} 
\item {\it Create labeled directories with data and plots for analysis and post-post-processing};
\item {\it Erase temporary data}. 
\end{itemize}
\end{itemize}}
\vspace{0.5cm}
\end{mdframed}}

\textcolor{black}{Also an extra automate tool was developed to analyze and visualize the whole set of results,  presented in the next section.
The features of the auxiliary script written in Python are:}
\begin{mdframed}
\textcolor{black}{
\begin{itemize}
  \item {\sc Input}: {\it Read two tables of temporal windows for all ICs};   
  \item {\sc Task One}:
     \begin{itemize} [label=\ding{212}]
         \item  For each IC:
            \begin{itemize} [label=$\checkmark$]
               \item {\it Prepare data: first subtraction and fit};
               \item {\it Calculate the putative QNM imaginary frequency};
               \item {\sc Output}: {\it Append each imaginary frequency};
             \end{itemize}
         \item {\it Generate and save a table.}
     \end{itemize} 
   \item {\sc Task two}:
      \begin{itemize} [label=\ding{212}]
         \item  For each IC:
            \begin{itemize} [label=$\checkmark$]
               \item {\it Prepare data: second subtraction and interpolate values};
               \item {\sc Output}: {\it Write a vector};
               \item {\it Calculate the putative QNM complex frequency};
               \item {\sc Output}: {\it Append each complex frequency};
             \end{itemize}
          \item {\it Generate and save a target plot in the complex plane for the bracketed ICs within some tolerance.}  
       \end{itemize}  
\end{itemize}
}
\end{mdframed}
\textcolor{black}{The first and second subtractions are explained in the next section.} 

\textcolor{black}{\section{Numerical results}}
\begin{figure}
\includegraphics[width=0.46\textwidth]{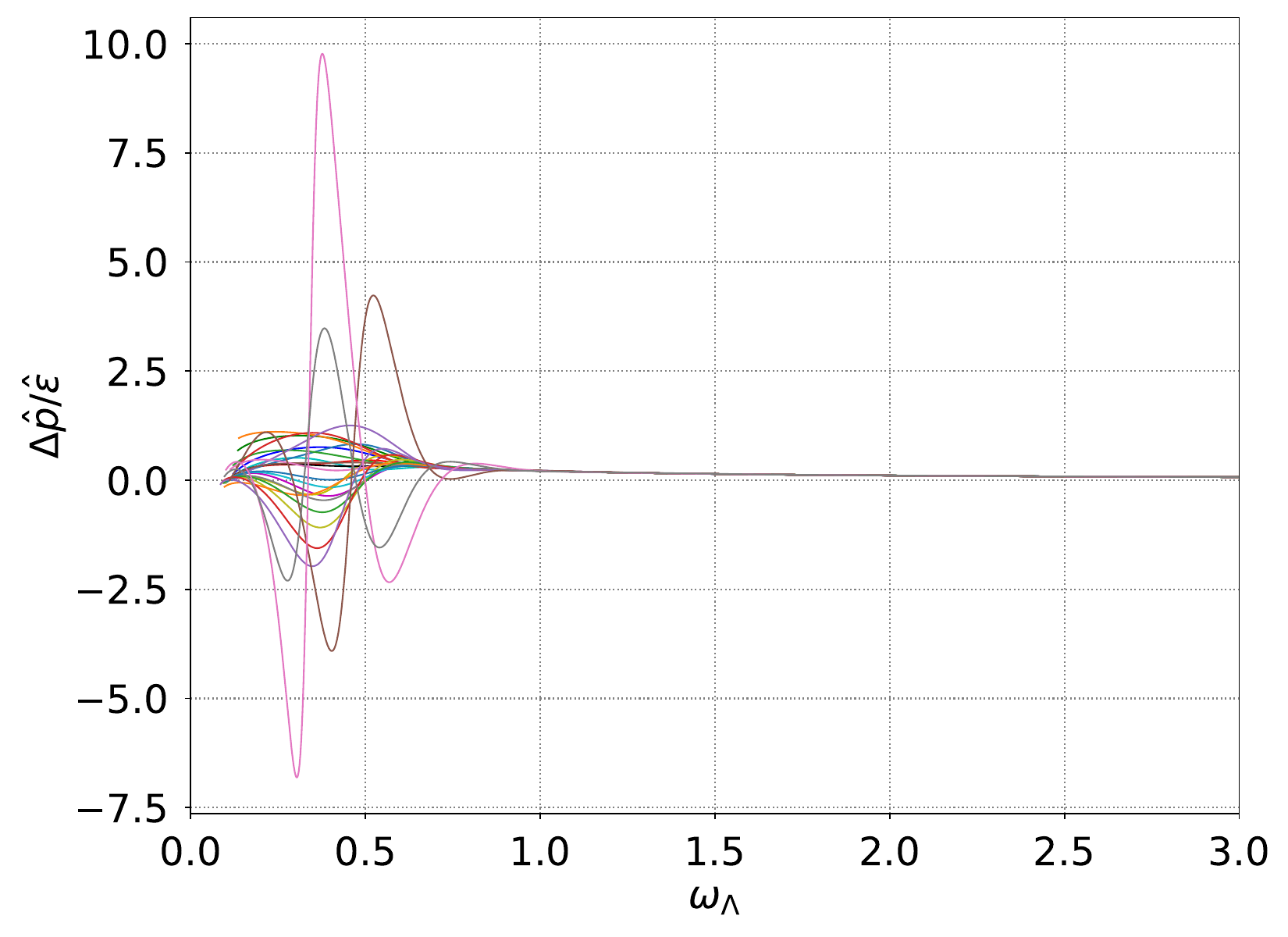}
\caption{Pressure anisotropy evolution for 25 IC as used in \cite{rnbdd21}, \cite{rbn22} and \cite{kaminski}.}
\label{fig:fig1}
\end{figure}
\begin{figure}
\includegraphics[width=0.46\textwidth]{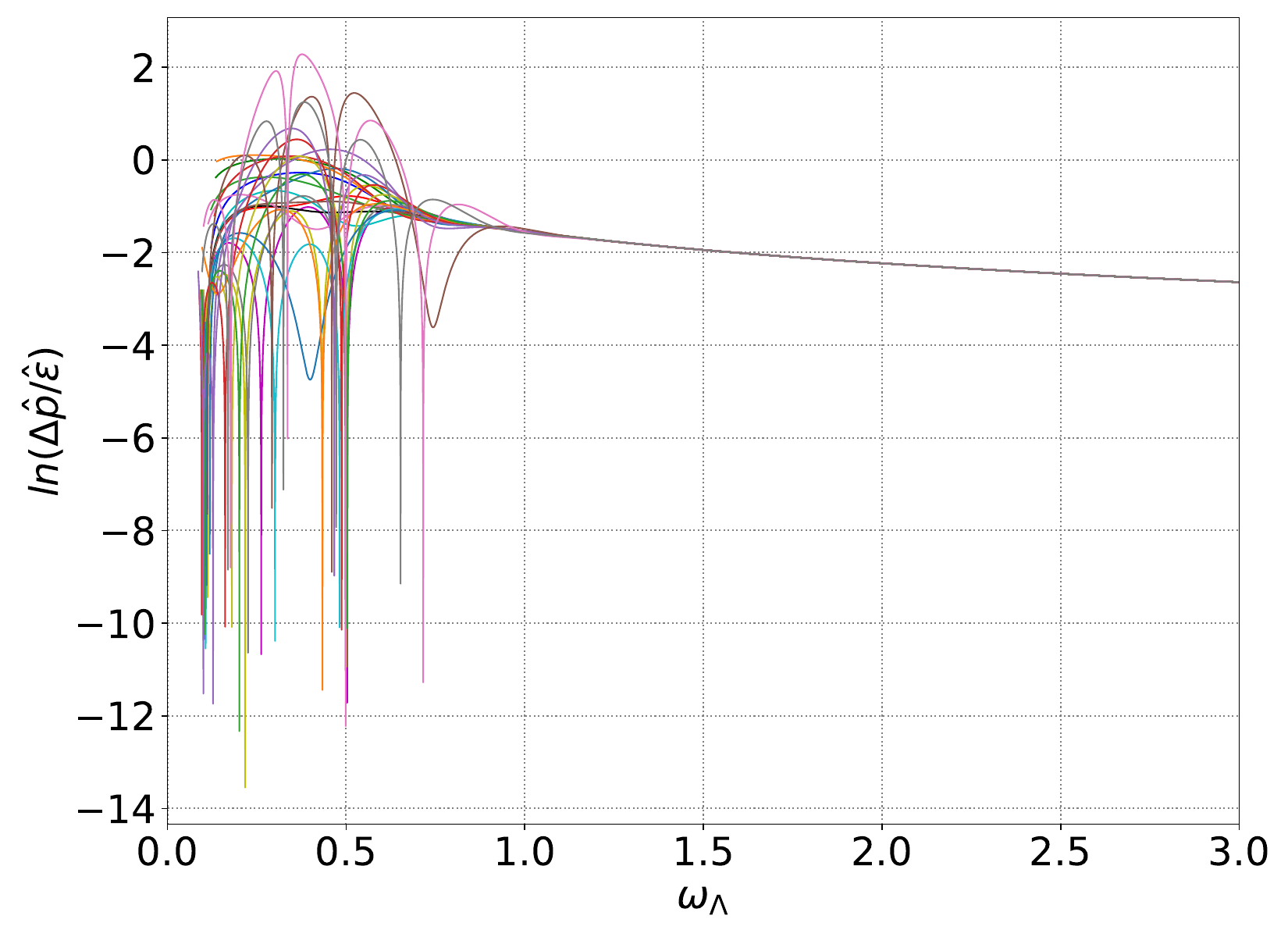}
\caption{Same as Fig. 1 in log scale.}
\label{fig:fig2}
\end{figure}
\begin{figure}
\includegraphics[width=0.5\textwidth]{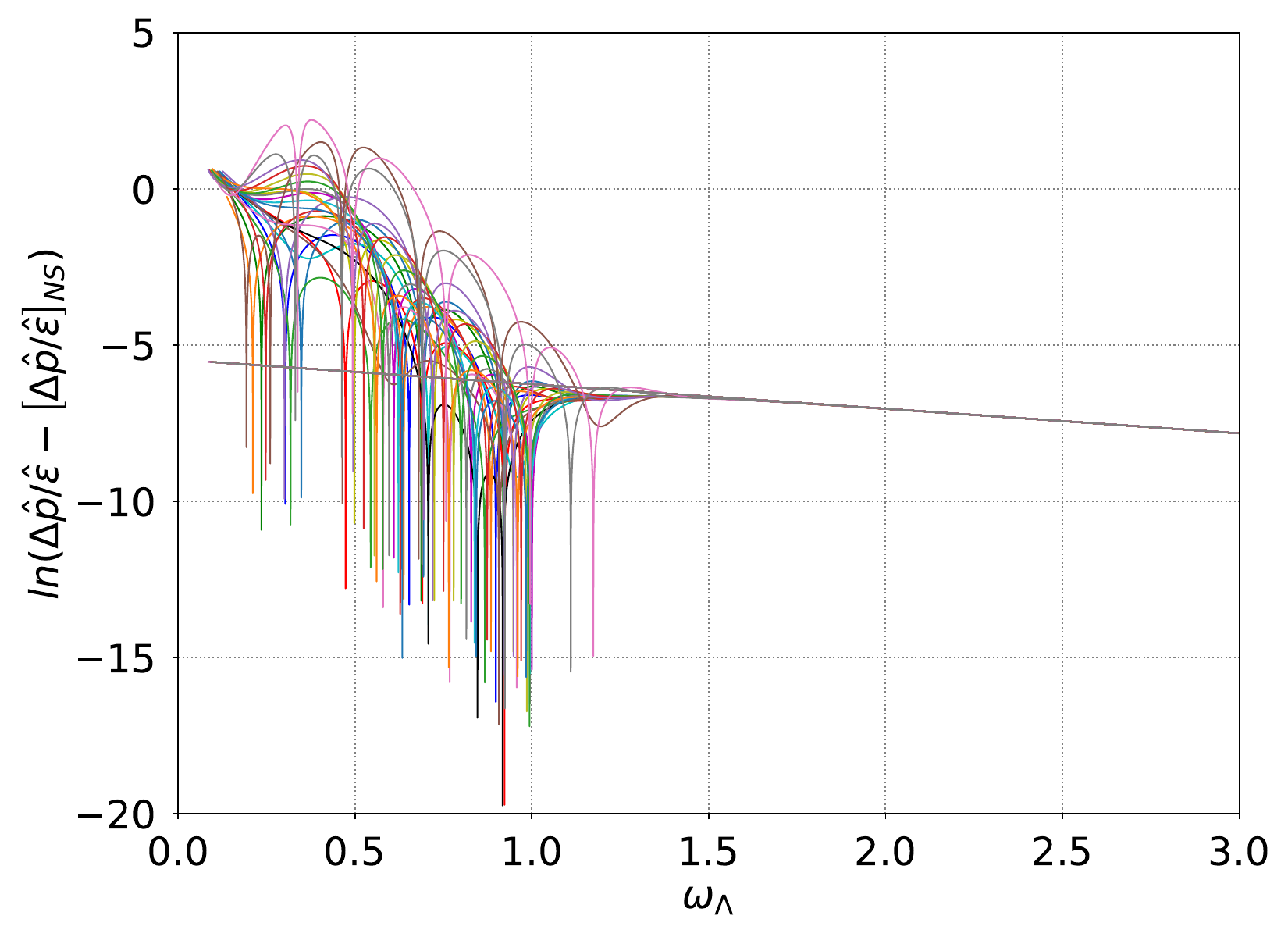}
\caption{Subtraction of the NS attractor to the pressure anisotropy $\Delta\hat p/\hat\epsilon-[\Delta\hat p/\hat\epsilon]_{NS}$ (shifted $c$). All the curves displayed fit with $\ln(a)-b\omega_\Lambda$ for late times.}
\label{fig:fig3}
\end{figure}
\begin{table}[h!]
\centering
\begin{tabular}{|c|c|c|c|}
\hline
\textbf{Model} & \textbf{NS} & \textbf{2nd Order} & \textbf{Borel} \\
\hline
1 & $0.786(4)$ & $1.73(2)$ & $2.07(2)$ \\
2 & $0.786(4)$ & $1.73(3)$ & $2.17(1)$ \\
3 & $0.786(4)$ & $1.73(2)$ & $1.97(2)$ \\
4 & $0.786(4)$ & $1.73(2)$ & $1.98(2)$ \\
5 & $0.786(3)$ & $1.73(2)$ & $1.83(2)$ \\
6 & $0.786(3)$ & $1.73(2)$ & $1.84(2)$ \\
7 & $0.786(4)$ & $1.73(2)$ & $1.96(2)$ \\
8 & $0.786(4)$ & $1.73(2)$ & $2.03(2)$ \\
9 & $0.787(5)$ & $1.73(3)$ & $1.91(2)$ \\
10 & $0.786(4)$ & $1.73(2)$ & $2.07(2)$ \\
11 & $0.786(4)$ & $1.73(3)$ & $2.15(2)$ \\
12 & $0.786(4)$ & $1.73(2)$ & $2.11(2)$ \\
13 & $0.786(4)$ & $1.73(2)$ & $1.97(2)$ \\
14 & $0.786(4)$ & $1.73(2)$ & $1.99(2)$ \\
15 & $0.786(3)$ & $1.73(2)$ & $1.80(2)$ \\
16 & $0.786(3)$ & $1.73(2)$ & $1.71(2)$ \\
17 & $0.786(3)$ & $1.73(2)$ & $1.86(2)$ \\
18 & $0.786(4)$ & $1.73(2)$ & $1.88(2)$ \\
19 & $0.786(3)$ & $1.73(2)$ & $1.78(2)$ \\
20 & $0.786(3)$ & $1.73(2)$ & $1.75(2)$ \\
21 & $0.786(3)$ & $1.73(2)$ & $1.65(2)$ \\
22 & $0.786(3)$ & $1.73(2)$ & $1.59(2)$ \\
23 & $0.786(4)$ & $1.73(2)$ & $1.96(2)$ \\
24 & $0.786(4)$ & $1.73(2)$ & $1.83(2)$ \\
25 & $0.786(3)$ & $1.73(2)$ & $1.78(2)$ \\
\hline
\end{tabular}
\caption{Decay parameter ($b$) fitted for each subtraction.}
\label{tab:tableI}
\end{table}

\begin{figure}
\includegraphics[width=0.5\textwidth]{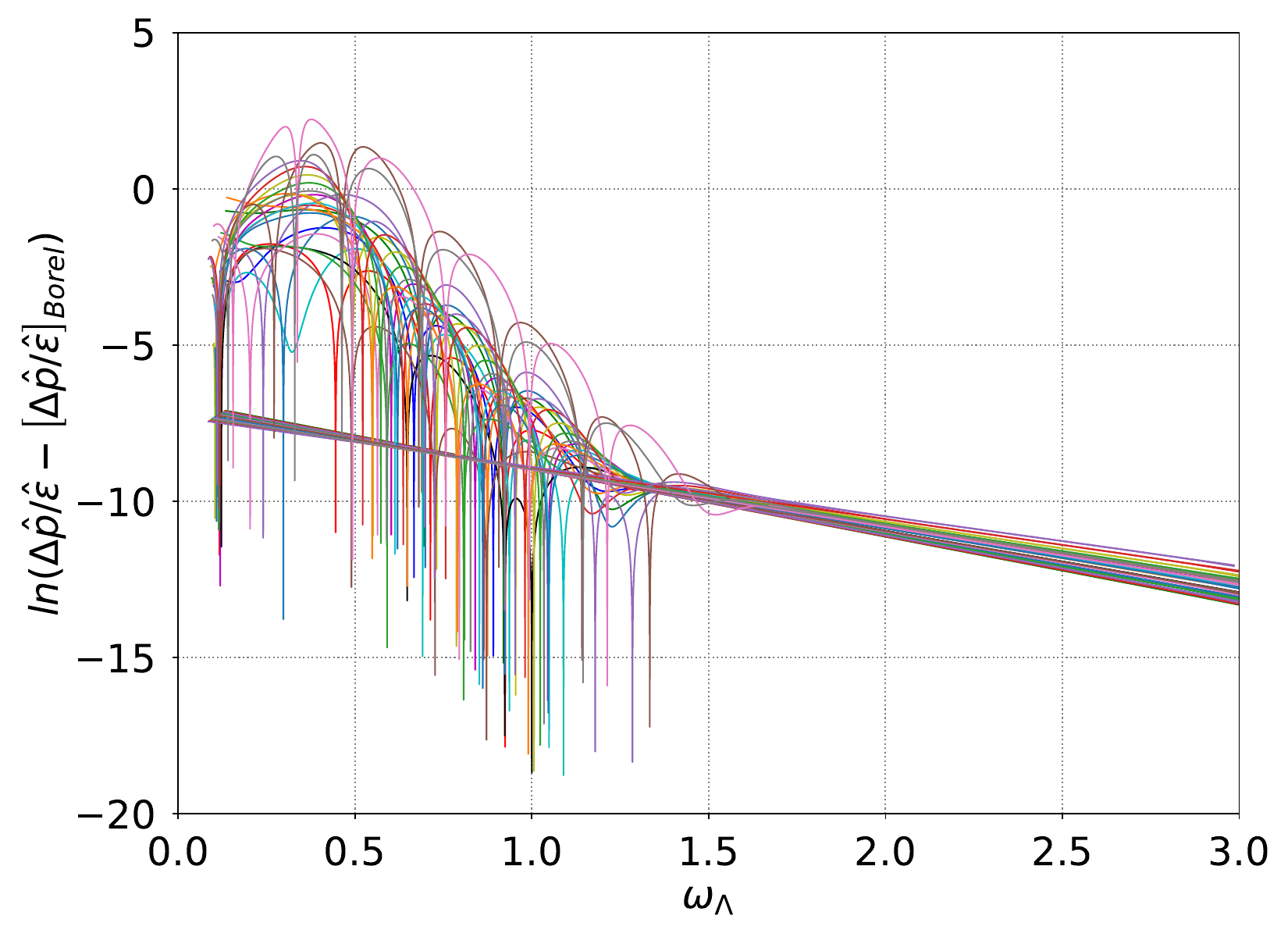}
\caption{Subtraction of the Borel attractor to the pressure anisotropy $\Delta\hat p/\hat\epsilon-[\Delta\hat p/\hat\epsilon]_{Borel}$ (shifted $c$). All the curves displayed fit with $\ln(a)-b\omega_\Lambda$ for late times.}
\label{fig:fig4}
\end{figure}
\begin{figure}
\includegraphics[width=0.5\textwidth]{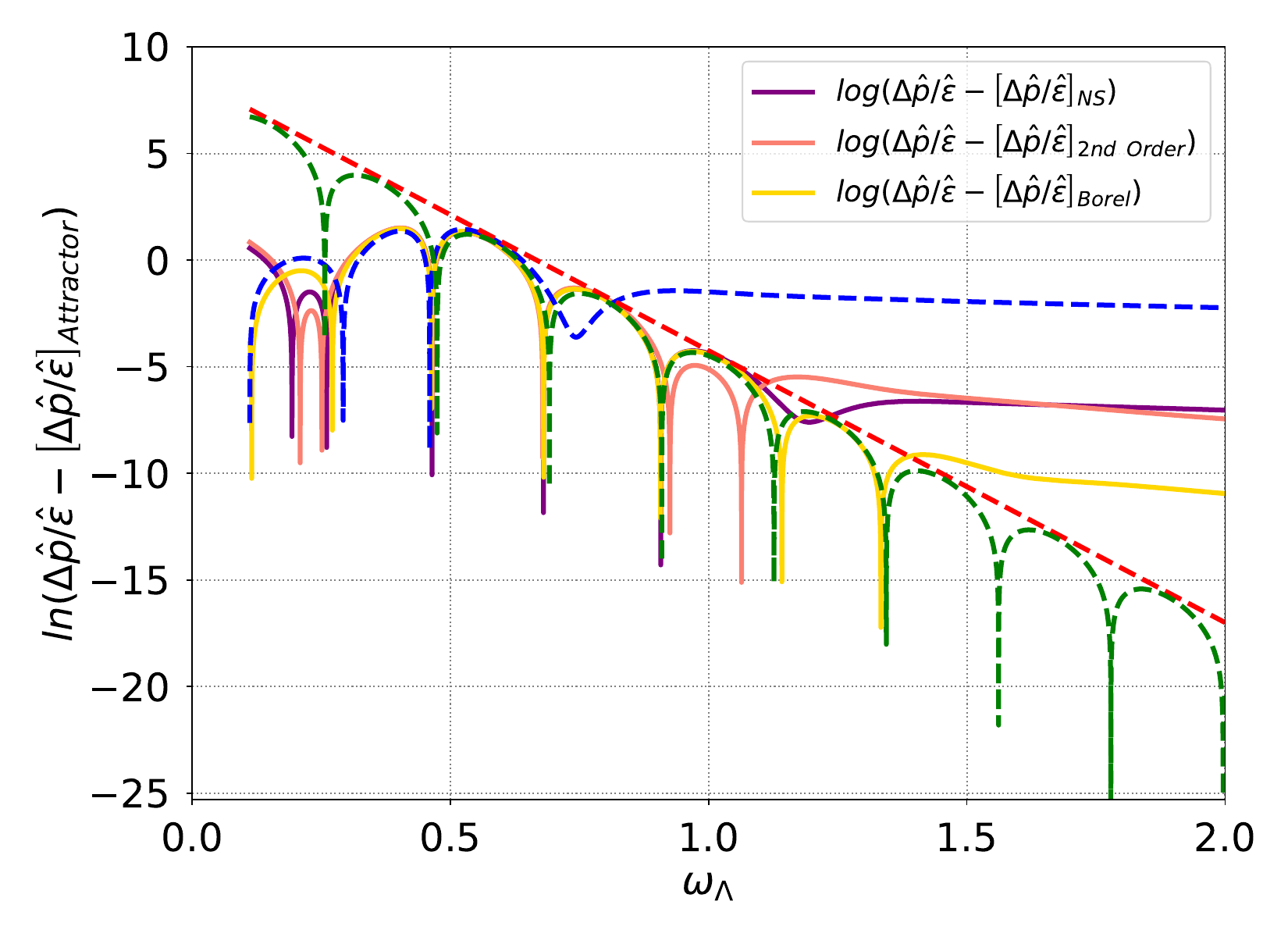}
\caption{Subtraction of attractors to the pressure anisotropy $\Delta\hat p/\hat\epsilon-[\Delta\hat p/\hat\epsilon]_{\text{Attractor}}$ for IC\#23: NS (violet), Second order (salmon), Borel (yellow). They are  compared with a modeled (guessed) QNM with complex frequency $14.45-i12.75$, shifted to fit visually (green dashed curve); the top line (red dashed line) is just a reference for the decay rate. The dashed curve in blue is the pressure anisotropy without subtraction.}
\label{fig:fig5}
\end{figure}
\begin{figure}
\includegraphics[width=0.5\textwidth]{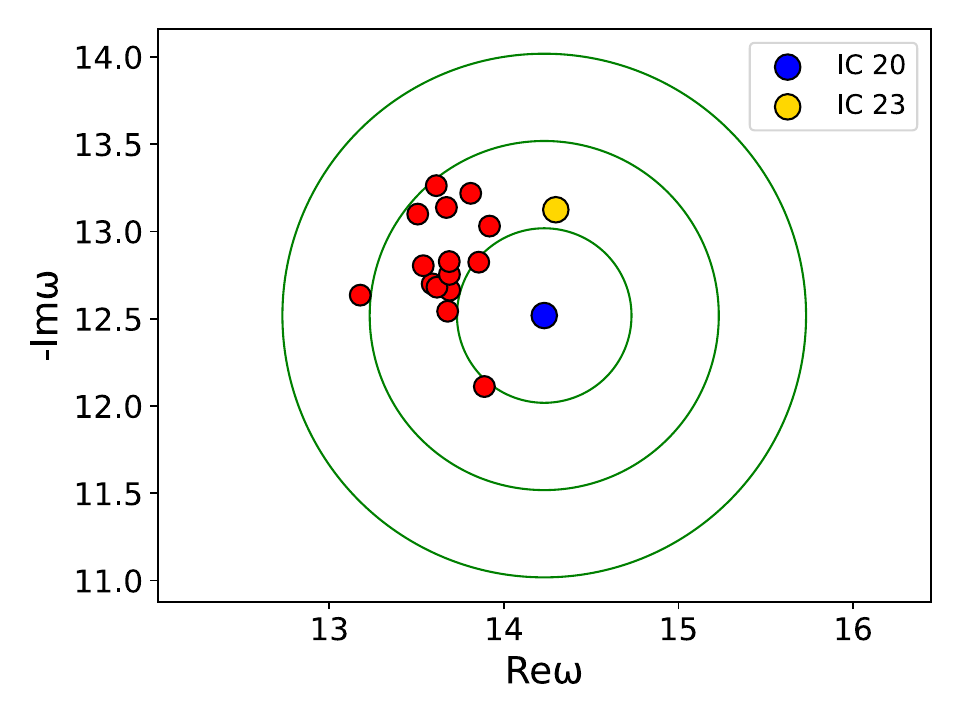}
\caption{Complex frequencies for the Borel subtraction, and for a number of IC computed with {\sc Harminv}. In this target plot the blue center choice is for IC\#20, and the gold circle is for the adjusted frequency for IC\#23. Other (red) IC are: 3-6, 12, 14-25 ($\approx$ 70\% of the IC set).}
\label{fig:fig6}
\end{figure}
\textcolor{black}{In the following sequence of results we begin by analyzing the pressure anisotropy as reported in \cite{rnbdd21}, \cite{rbn22} and \cite{kaminski} (see figure 1), for the set of 25 IC, now for late times. Figure 2 displays the same results in log scale. Clearly for late times all IC decay apparently to the same tail. We subtract the attractors (\ref{attractors}), one at a time, to determine the law of the decay rate for late times (this simple operation is justified in section V). Doing the subtraction, the best trial fit function for the three attractors and for all IC with $\omega_\Lambda \in [2.0,3.5]$, was ($ae^{-b\omega_\Lambda} +c$), not ($ae^{-b\omega_\Lambda}$). All IC have the same decay ($b$) for the subtracted NS and second order attractors, as shown in Table I, within the indicated numerical error. Figures 3 and 4 display the subtraction for NS and Borel. These results are expected, discussed and explained in the next section. This is the first subtraction as commented in section III.}

\textcolor{black}{What turns out to be more interesting is the revelation of transient oscillations with period and decay (numerically close), for a significantly number of IC. That behavior is more persistent in time for the Borel attractor subtraction (see figure 4). To show that clearly we isolate the IC\#23, for instance. Figure 5 displays the subtracted pressure anisotropy for this IC, compared with the non-subtracted pressure anisotropy and with a crudely guessed QNM.  
In consequence we define a second task, that is, the systematic extraction of complex frequencies in specific windows of time for each IC and for the most persistent signal, that is, for the Borel subtraction. This is the second subtraction as commented in section III. To analyze the selected data we use {\sc Harminv} \cite{harminv}, a well known and robust software to extract QNMs. This in turn requires the interpolation for equal time-step in $\omega_\Lambda$, for which we choose the standard Spline method. The results are displayed in figure 6. Thus, we have collected numerical evidence of a collective behavior for the studied set of IC. But, are these really QNMs? If so, are they related to other theoretically expected QNM? In the following section we discuss our results.}
\vspace{0.2cm}
\textcolor{black}{\section{Discussion}}
\textcolor{black}{It was a matter of logical (and simple) operation subtracts an attractor to capture the hidden  decaying oscillations, like the QNM. In this sense we subtracted hydrodynamical modes, homogenizing the Bjorken flux. Doing that, late time tail and transient QNM emerge and characterized numerically. Subtracting an equilibrium function, an attractor, it is reasonable to think that the best one for late time is the NS attractor. This leads us to the approximate value for the decay $b\approx 0.79$ for all IC.  
This can be interpreted as an imaginary frequency in the context of QNM analysis. Then it was manifest, before the transition to a tail, transient QNM signals with complex frequencies. Visually, the decay rate and the period appeared to be similar for a good number of evolved IC, at least in some window (see Figs. 3 and 4). To visualize a representative evolution we select the IC\#23 evolution. It was clear enough that the subtraction of the Borel attractor revealed a more persistent QNM that for other attractors (see Fig. 5). As displayed in Fig. 6, they are not exactly the lowest QNM but they are significantly close to.}

\textcolor{black}{Our results motivated us to establish connections with the results of other authors (including own work). Without surprise, it is good to extract numerical evidence of the expected lowest QNM for the Bjorken flux, from a rapid calculation and a relatively low precision code. In \cite{jp06}, Yanik and Peschanski obtain the QNM for a scalar perturbation of a background 5-d planar black hole geometry, which is $\omega_{static}/\pi T= 3.1194-2.74667i,$ calculated previously by Starinets \cite{s02} as a fundamental QNM near extremal black branes dual to strongly coupled $\mathcal{N}=4$ SYM plasma at finite temperature. Using the dual gravity description in \cite{hjw13}, the authors calculate numerically the form of the stress tensor for a boost-invariant flow in a hydrodynamic expansion up to terms with 240 derivatives. They identify the leading singularity in the Borel transform of the hydrodynamic energy density with the lowest nonhydrodynamic excitation corresponding to a ‘nonhydrodynamic’ quasinormal mode on the gravity side. Explicit gravity calculation for the lowest mode yield 
$\omega_{qnm}=3.1195-2.7467i,$ which agrees with the frequency of the lowest nonhydrodynamic scalar quasinormal mode as calculated in \cite{jp06}. Then they reproduce numerically that mode from the large order behavior of the hydrodynamic series\footnote{The working precision in \cite{hjw13} is $10^{-100}$.}, obtaining $\omega_{\text{Borel}}=3.1193-2.7471 i$. In \cite{rb24} for homogeneous isotropization we obtain for the SYM model the lowest QNM for the quintuplet channel $\omega_{HI}/T \approx 9.8-8.6i$ (in agree with \cite{crn17}), which is equal to (except by a scale factor of $T$ and working precision) the other authors QNM complex frequency, $\approx\pi(3.12-2.75)\approx 9.8-8.6i$. Now, if we consider the ratio of the Borel frequency 
$-\Re(\omega_{\text{Borel}})/\Im(\omega_{\text{Borel}})\approx 1.1345$ and compare it with the same ratio for our centered complex frequency (for IC\#20), calculated with {\sc Harminv} (see Fig, 6) in post-post-processing, we get $\approx 1.1369$, which is close enough to extract a scale factor for each evolved IC. For our centered complex frequency in Fig, 6 (IC\#20) the scale factor is $\approx 0.219$. Thus, $0.219\times(14.231-12.518i)\approx 3.12-2.74i$, in agree with the expected values, within numerical errors.} 

\textcolor{black}{Shedding non-equilibrium (Borel's regime), the system is driven to equilibrium, characterized by transient QNM. For the most of IC is revealed that the corresponding QNM is close to the expected lowest QNM. Then, any of the considered IC evolves to the stable (pure imaginary) QNM (NS's regime). If there exits a second order regime it is intermediary.}

\textcolor{black}{It seems that the function of the attractor is to drive the system to the threshold of hydrodynamization from the initial state far-from-equilibrium. The system chase the QNM, particularly when the energy conditions are being (or close to be) violated, looking for the exit and evolve to equilibrium. This could explain the formation of plateaus for entropy far-from-equilibrium. In the way to equilibrium the QNM do not persist for late times, where a pure decay dominates, without a stairway formation. It is interesting to observe that the plateaus for entropy are not in phase with the QNM, anticipating the violation of energy conditions as observed in previous works \cite{rnbdd21}, \cite{rbn22}, \cite{rb24}.}

\textcolor{black}{QNMs and tails are signatures of gravitational radiation from binary black holes (see \cite{buonano24} for a recent work, for example), in the astrophysical context. Chesler and Yaffe \cite{cy09}, in the context of horizon formation and far-from-equilibrium isotropization in a SYM plasma,  pointed that non-monotonicity is unsurprising because the late time response is dominated by the lowest quasi-normal $(\pm9.8-8.7i)T$ \cite{s02}. They also point out, in the same context, the presence of infalling gravitational radiation. Thus, the tails for late time in this work is not a surprise, and neither a (transient) lowest QNM.
The computational procedure, as implemented in this work, can be extended to the holographic model 1RCBH undergoing a Bjorken flux \cite{rb22}. Although for this last model the Borel attractor is not known, as far as we know, the subtraction of the NS attractor catchs a significant part of the QNM before the final relaxation (decay).}

\textcolor{black}{Finally, we think that this work was a nice example of deep human learning, and could be interesting for new practitioners in the field.}

\textcolor{black}{\acknowledgements}
\textcolor{black}{GG and WB thanks FAPESP, Scientific Initiation Program, under grant 2023/07953-5. WB thanks FAPESP, Research Projects Program, under grant 2022/02503-9 and acknowledge the financial support by National Council for Scientific and Technological Development (CNPq) under grant number 407162/2023-2. Also we thank to Nairy Villarreal for comments about the original version of this work. The authors thank to the {\it Central de Computação Multiusuário} (CCM) at UFABC, for support using the clusters {\it Titânio} and {\it Carbono}.}  
\end{document}